\documentclass{jpsj3}  %
\usepackage{txfonts}

  \title{
Spontaneous Magnetic Field and Local Density of States 
near a Time-Reversal Symmetry Broken Surface State of YBCO 
}
 
\author{Kazuhiro Kuboki \thanks{kuboki@kobe-u.ac.jp}}
\inst{Department of Physics, Kobe University, Kobe 657-8501, Japan
}  
 
\abst{    
We study theoretically the spatial distribution of spontaneous magnetic 
fields near a (110) surface of cuprate high-$T_C$ superconductor 
YBCO by treating a bilayer $t-J$ model 
with the Bogoliubov-de Gennes method. 
Near a (110) surface where $d_{x^2-y^2}$-wave 
superconductivity is strongly suppressed, 
flux phase order occurs locally leading to a time-reversal 
symmetry broken surface state.  
Since the spontaneous currents flowing in the flux phase 
in two layers are antiparallele, 
resulting spontaneous magnetic fields are quite small. 
They exist in a narrow region inside the superconductor, 
and essentially vanish outside irrespective of the temperature.  
This result, together with the splitting of zero-energy 
peak of the local density of states, can provide 
consistent explanations for several experiments on YBCO. 
}
 
\begin{document}
\maketitle

\newpage

Spontaneous breaking of time-reversal symmetry  (${\cal T}$) near 
a surface or grain boundary of superconductors attracts much 
attention, because it is closely 
related to the symmetry of superconducting (SC) states and the 
mechanism of superconductivity. 
In the case of cuprate high-$T_C$ superconductors,  
splitting of zero-bias conductance peak in 
YBCO/insulator/Cu junction was observed, \cite{Coving} 
and it has been regarded as a sign of spontaneous violation of 
time-reversal symmetry.\cite{SigRev} 
This experimental finding has been theoretically interpreted as 
due to the appearance of a second SC order parameter (OP) 
that has symmetry different from that in the bulk 
($d_{x^2-y^2}$ wave). 
Since the $d_{x^2-y^2}$-wave SC ($d$SC) state is strongly suppressed 
near a (110) surface,  ${\cal T}$-breaking SC states with 
$(d \pm is)$-wave symmetry may occur near the surface.  
\cite{Matsu1,Matsu2,Fogel,1}
In these states, spontaneous currents and magnetic fields are 
expected to arise, but 
their existence is still controversial.\cite{Carmi,msr} 
(Also, the splitting of the peak is not always observed in 
experiments in similar situations.\cite{4b})

In order to explain these phenomena, 
H\r{a}kansson et al.\cite{Fogel2} considered the edge state of 
a $d$-wave superconductor where neighboring vortices have 
opposite current circulation. This state breaks ${\cal T}$,  
and gives rise to split peaks in the local density of states (LDOS) 
and a relatively small spontaneous magnetic field. 
Potter and Lee\cite{Lee2} discussed ferromagnetism induced 
at the edge of $d$-wave superconductor 
that also leads to the splitting of the zero-energy peak.

The present author studied (110) surface states of YBCO 
that has two CuO$_2$ planes in a unit cell,  
using a bilayer $t-J$ model and the Bogoliubov-de Gennes (BdG) 
method.\cite{KK3,KK4}  
Near the (110) surface where the $d_{x^2-y^2}$-wave SC state 
is strongly suppressed, a flux phase appears locally leading to a
${\cal T}$-violating surface state. 
The flux phase is a metastable mean-field solution to the $t-J$ model 
in which staggered currents flow and  a flux penetrates a plaquette 
on a square lattice.\cite{Affleck} 
(The $d$-density wave state, which was introduced
in a different context, have similar properties.\cite{Chakra}) 
Mean-field calculations \cite{Zhang,Ubb,Hamada,Zhao,Bejas,KK1}
and variational Monte Carlo (VMC) 
study\cite{Yoko1,TKLee,Ivanov} 
have shown that free energy of the flux state is higher than that of 
$d$SC state except very near half filling, 
so that it is only a next-to-leading (metastable) 
state in uniform systems.
(Inclusion of nearest-neighbor repulsive interactions to the model 
may lead to the appearance of the flux phase for small 
doping rates.\cite{Zhou})
VMC calculations for the Hubbard model have shown 
that coexistence of flux phase and superconductivity is 
not possible in uniform systems,\cite{Yoko2} 
in agreement with the results for the $t-J$ model.

When the shape of the Fermi surface is favorable to 
the $d$SC state, 
other kind of ordered states that has the same symmetry 
may have free energy close to that of the former. 
This is because self-consistency equations in both states have the 
same form factor $(\cos k_x - \cos k_y)^2$, 
where ${\bf k} = (k_x, k_y)$ is a wave vector in the Brillouin zone, 
and it is large on the Fermi surface. 
On the contrary, the $s$-wave SC ($s$SC) state 
has higher free energy, since the corresponding form factor 
$(\cos k_x +\cos k_y)^2$ is small on the same 
Fermi surface.\cite{KK2001}
(For an onsite pairing SC state, the form factor is a 
${\bf k}$-independent constant, 
but this pairing is not allowed in the strongly correlated systems.)
The flux phase has $d_{x^2-y^2}$-wave symmetry, so that 
its free energy is close to that of $d$SC state compared with  
the $s$SC state.  
Thus, once the $d$SC order is suppressed, 
the flux phase can be a leading candidate of the local ordered state 
near the surface.

When the flux phase arises, spontaneous currents 
flow along the surface as in $(d \pm is)$-wave SC state. 
However, the current in this case is oscillating as a function 
of the distance from the surface. Thus the spontaneous 
magnetic field generated by this current is expected to be smaller 
compared with that in the surface $(d \pm is)$-wave SC state.
Moreover in the bilayer $t-J$ model that describes the low-energy
electronic states of YBCO, the directions of both fluxes  
and currents in the flux phase are opposite 
in two layers\cite{Zhao,KK3} 
This should further reduce the absolute values of spontaneous 
magnetic fields. 
We have investigated this problem 
using the Ginzburg-Landau (GL) theory derived from the 
$t-J$ model,\cite{KKGLFL,KKGLFL2} 
and showed that magnetic fields are actually very small. 
They almost vanish outside the superconductor, although 
they are finite inside the sample 
within a narrow region of SC coherence length, $\xi_d$, 
from the surface. 
However, the temperature ($T$) region in which the GL theory 
is quantitatively reliable is limited to that near $T_C$, i.e.,  
the SC transition temperature. 

In this letter, we study the spatial distribution of the spontaneous 
magnetic field near the (110) surface of YBCO, using the BdG method 
applied to the bilayer $t-J$ model.
We examine their temperature dependence, and 
show that even at lowest $T$ the spontaneous magnetic fields are 
measurable only in a narrow region inside the superconductor 
(with a width of the order of $\xi_d$),  
and in a quite narrow region outside.  
By measuring these magnetic field distributions experimentally, 
it will be possible to decide whether the present theory 
correctly describes the surface state of YBCO. 
We also study the temperature dependence of the 
LDOS near the surface,  which can also be tested experimentally.

We treat the  bilayer $t-J$ model whose Hamiltonian is given by 
$H = H_1 + H_2 + H_T$ with 
\begin{eqnarray}
\displaystyle 
H_i =&\displaystyle -\sum_{j,\ell,\sigma} 
t_{j\ell} e^{i\phi_{j,\ell}} 
{\tilde c}^{(i)\dagger}_{j\sigma} {\tilde c}^{(i)}_{\ell\sigma}
 +J\sum_{\langle j,\ell\rangle} {\bf S}^{(i)}_j\cdot {\bf S}^{(i)}_\ell  
 \ \ \  (i=1,2)\\
 H_\perp =&\displaystyle -\sum_{j,\ell,\sigma} t_{j\ell}^\perp 
\Big( {\tilde c}^{(1)\dagger}_{j\sigma} {\tilde c}^{(2)}_{l\sigma} + h.c.\Big)
 +J_\perp\sum_j {\bf S}^{(1)}_j\cdot {\bf S}^{(2)}_j, 
\end{eqnarray}
where $H_1$ and $H_2$ are the Hamiltonian for each layer
(with a square lattice)
and $H_T$ describes interlayer couplings. 
${\tilde c}_{j\sigma}^{(i)}$ is the electron operator at a site 
$j$ on the $i$-th plane with spin $\sigma$ in Fock space without 
double occupancy, and we treat this condition 
using the slave-boson (SB) method.\cite{OF,LNW}  
Here ${\bf S}_j$ denotes the spin operator.
(For the detail of SB mean-field calculation, see Ref. 13.)
The in-plane transfer integrals  $t_{j\ell}$ are finite for the first-  ($t$), 
second-  ($t'$), and third-nearest-neighbor bonds ($t''$), or zero otherwise.  
$J$ $(J_\perp)$ is the inplane (interplane) antiferromagnetic superexchange
interaction, and $\langle j,\ell \rangle$ denotes nearest-neighbor bonds.
The interplane transfer integrals $t^\perp_{j\ell}$ are chosen to reproduce 
the dispersion in $k$ space,\cite{Andersen} 
$t^\perp_k =  -t^\perp_0 (\cos k_x - \cos k_y)^2$, namely, 
"on-site" ($t^\perp_0$), second- ($t^\perp_2 = -t^\perp_0/2$) , 
and third-nearest-nearest-neighbor bonds ($t^\perp_3 = t^\perp_0/4$) 
are taken into account.  
The magnetic field is taken into account using the Peierls phase 
$\phi_{j,\ell} \equiv \frac{\pi}{\phi_0} \int_j^\ell {\bf A}\cdot d{\bf l}$, 
with ${\bf A}$ and $\phi_0 = \frac{h}{2e}$ being the vector potential 
and  flux quantum, respectively. 
Following Ref. 33,
we take $t/J=2.5 \ (J=0.1$eV), $t'/t=-0.3$, $t''/t=0.15$, 
$t^\perp_0/t =0.15$, and $J_\perp/J = 0.1$.
These parameters were chosen to reproduce experimental 
results for YBCO.
We take $\delta=0.15$ for the doping rate throughout in this work.
The results for other values of $\delta$ are qualitatively the same; 
for larger value of $\delta$, Im $\chi$ becomes smaller and so the
absolute value of $B$. 
In mean-field calculations for uniform systems, 
the flux phase with opposite directions of fluxes in two layers 
persists only up to $\delta \sim 0.15$.\cite{KK3}
However in the BdG calculation it can occur up tp $\delta \sim 0.3$, 
because the incommensurate solution that is not taken into account 
in the former can arise, and it has free energy lower than that of the
flux phase with the same direction of fluxes.\cite{KK4}

We consider a system with a (110) surface, and denote the direction 
perpendicular (parallel) to the surface as $x$ ($y$), where 
the region $x > 0$ ($x < 0$) is a superconductor (vacuum). 
We assume that the system is uniform along the $y$ direction,  
and the periodic boundary condition is used for this direction. 
For the system size, $N_x=200$ and $N_y=100$ are used.
Following the procedure presented in Refs. 13 and 34, 
we perform BdG calculations to study spatial variations 
of the $d$SC ($\Delta_d$) and $s$SC ($\Delta_s$) 
OPs, and the bond OP ($\chi$). 
The imaginary part of $\chi$ corresponds to the OP for 
the flux phase.

In Fig.1, we show the spatial variations of the OP for the flux phase, 
Im $\chi$, for several temperatures.
Surface flux phase appears below $T=0.19T_C$ ($T_C = 0.1164J$),
and  Im $\chi$ becomes larger as $T$ is lowered. 
The spontaneous current along the surface in the $i$-th layer, 
$J_y^{(i)}$, is proportional to Im $\chi^{(i)}$ as\cite{KK2,KK4} 
\begin{equation}
\displaystyle 
J_y^{(i)}(x) = \frac{\sqrt{2} \pi t \delta}{\phi_0} 
{\rm Im} \ \chi^{(i)}(x) \ \ \ \ \ \  (i=1,2). 
\end{equation} 
The currents in two layers are antiparallel: 
$J_y^{(1)}(x) = -J_y^{(2)}(x)$.

\begin{figure}[htb]
\begin{center}
\includegraphics[width=7.0cm,clip]{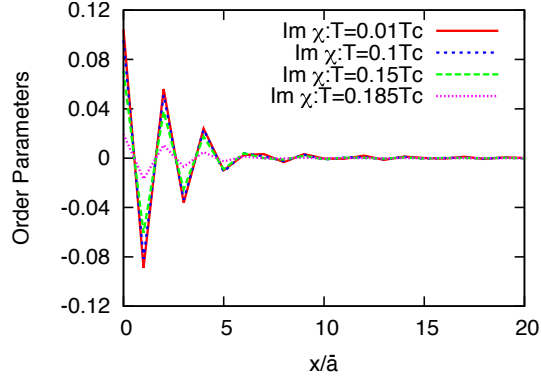}
\caption{(Color online) Spatial variations of the imaginary part   
of the bond order parameter, Im $\chi$, on the first layer ($i=1$).
Note that $\chi$ is nondimensional.}
\end{center}
\end{figure}

Next we calculate spontaneous magnetic fields 
generated by these currents.
We model the actual system by considering the infinite stacking 
of  bilayer systems.
Assuming the tetragonal structure, $c$-axis lattice constant 
and the distance between a bilayer are taken to be $c=11.7$ \AA \
and $c_1=3.4$ \AA, respectively to represent the structure of YBCO. 
We take the origin of the $z$ axis at the center of a bilayer.
(Distance between neighbor currents in a plane is 
${\bar a} \equiv a/\sqrt{2}, $\cite{KK2,KK4} where $a=3.8$ \AA \
is the $a$-axis lattice constant.)
Magnetic fields ${\bf B}$ are estimated by using the Biot-Savart law 
\begin{eqnarray}
\displaystyle 
B_x(x,z) =  & \displaystyle \frac{\mu_0}{2\pi} 
\sum_{j=1}^{N_x-1}\sum_{n=-\infty}^{\infty} 
\Bigg[\frac{(z-z^{(+)}_n)J^{(2)}_y(x_j)}{(x-x_j)^2+(z-z^{(+)}_n)^2}
+ \frac{(z-z^{(-)}_n)J^{(1)}_y(x_j)}{(x-x_j)^2+(z-z^{(-)}_n)^2}\Bigg]
\\
B_z(x,z) =  & \displaystyle -\frac{\mu_0}{2\pi} 
\sum_{j=1}^{N_x-1}\sum_{n=-\infty}^{\infty} 
\Bigg[\frac{(x-x_j)J^{(2)}_y(x_j)}{(x-x_j)^2+(z-z^{(+)}_n)^2}
+ \frac{(x-x_j)J^{(1)}_y(x_j)}{(x-x_j)^2+(z-z^{(-)}_n)^2}\Bigg], 
\end{eqnarray}
where $x_j =( j-1/2){\bar a}$ and $z^{(\pm)}_n = cn \pm c_1/2$. 
The expression for $B_z$ can be simplified as 
\begin{equation}
\displaystyle 
B_z(x,z) = \frac{\mu_0}{2\pi c} \sum_{j=1}^{N_x-1}
\big[G^{(+)}(x,z,j) - G^{(-)}(x,z,j)\big] J^{(1)}_y(x_j),
\end{equation}
where 
\begin{equation}
\displaystyle 
G^{(\pm)}(x,z,j) = \frac{\pi \sinh\big(\frac{2\pi}{c} (x-x_j)\big)}
{\cosh\big(\frac{2\pi}{c} (x-x_j)\big) 
- \cos\big(\frac{2\pi}{c}(z\mp\frac{c_1}{2})\big)}.
\end{equation}
In this approach the screening effects in superconductors 
are neglected, so that the results should be taken as upper limits 
of the absolute values of $B$.

We show the $x$ dependence of the magnetic fields 
inside the superconductor for two choices of $z$ in Fig.2.
Here the temperature is $T = 0.01T_C$. 
It is seen that both in-plane ($B_x$) and vertical ($B_z$) 
components occur near the surface 
and they are oscillating as functions of $x$.
(For $z=0$, $B_z$ vanishes due to symmetry.)
The important feature is that the magnetic fields decay 
quite rapidly as functions of $x$. 
This is because contributions from antiparallel currents 
cancel each other at a site where $J_y \sim 0$. 
Actually, we can see from Eqs. (6) and (7) 
that the typical length scale of the decay of $B_z$ is $c/2\pi$. 
The amplitude of the magnetic field is large near the surface 
where $J_y$ is finite. 
These large magnetic fields existing in a narrow region 
of $x \lesssim \xi_d$ may be detected by experimental 
approaches such as $\mu$SR or polarized neutron scattering, 
although the values of $B$ will be reduced if one 
treats the screening effect correctly. 
(Surface roughness, which is inevitable in real materrals, 
would also reduce the value of $B$.) 
If $J_y(x)$ does not oscillate and has definite sign, 
$B$ will be finite for $x \lesssim \lambda$  
($\lambda$ being the penetration depth), 
even when we treat the screening effect properly. 
%

\begin{figure}[htb]
\begin{center}
\includegraphics[width=7.0cm,clip]{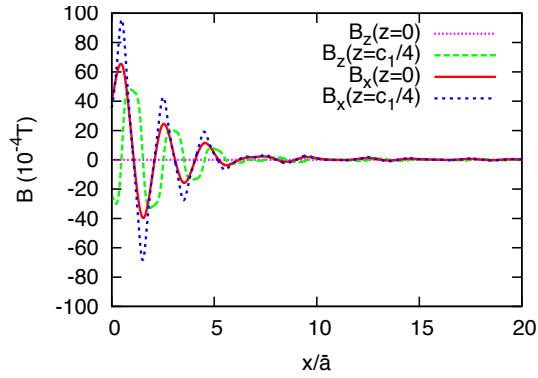}
\caption{(Color online) Spatial variations of $B_x$ and $B_z$ 
as functions of the distance from  the surface, $x$, 
for $z=0$ and $c_1/4$. 
The temperature is $T=0.01T_C$.} 
\end{center}
\end{figure}

In Fig.3, $B_x$ and $B_z$ are presented as functions of $z$ 
for two choices of $x$ ($<0$), i.e., outside the superconductor. 
Here the points $z=0$ and $\pm c$ are the center of a bilayer 
in neighboring unit cells. 
Outside the sample $B$ decays quickly in a manner similar to 
the case of $x \gtrsim \xi_d$ (i.e., inside the superconductor) 
due to cancellation among 
contributions from antiparallel currents.
Although the values of $B$ outside the superconductor 
are quite small, we may expect that the spin polarized 
scanning tunneling microscopy (STM) could measure 
the spatial distribution of the magnetic field.

\begin{figure}[htb]
\begin{center}
\includegraphics[width=7.0cm,clip]{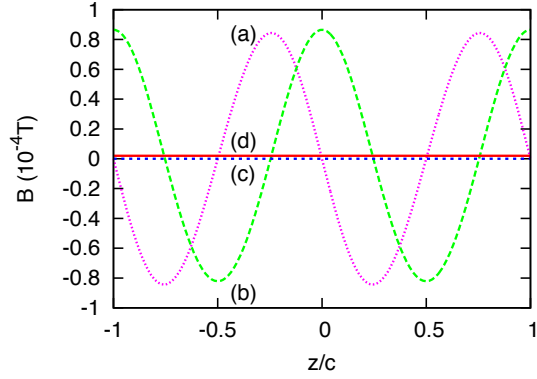}
\caption{(Color online) z dependence of $B_x$ and $B_z$ outside 
the superconductor for $T=0.01T_C$. 
(a) $B_z(x=-0.2c,z)$, (b) $B_x(x=-0.2c,z)$, 
(c) $B_z(x=-c,z)\times10$, and (d) $B_x(x=-c,z)\times10$.}
\end{center}
\end{figure}

In order to examine the temperature dependence of the magnetic field, 
we plot their variations along the $x$ ($z$) direction for 
$T = 0.185T_C$ in Fig. 4 (Fig. 5).
As the temperature is raised, the OP for the flux phase 
decreases as shown in Fig.1. Then the amplitudes of the 
magnetic fields are expected to decrease in proportion to it. 
This is actually the case as we can see by comparing 
Fig.4 (Fig.5) with Fig.2 (Fig.3). 

In the scenario to explain ${\cal T}$ violation using the 
second SCOP, e.g., $(d\pm is)$-wave SC states, spontaneous 
currents on different layers would be parallel, since 
Josephson coupling between layers would 
favor phase difference of the $s$SC OPs to be zero.
Thus ${\bf B}$ should be observable outside the sample
in this case.

\begin{figure}[htb]
\begin{center}
\includegraphics[width=7.0cm,clip]{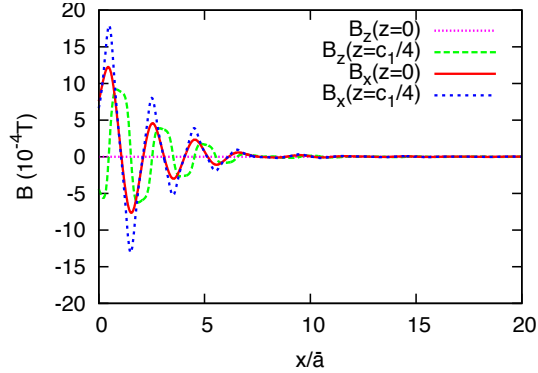}
\caption{(Color online) Spatial variations of $B_x$ and $B_z$ 
as functions of the distance from the surface, $x$, for 
$z=0$ and $c_1/4$.
The temperature is $T=0.185T_C$.} 
\end{center}
\end{figure}

\begin{figure}[htb]
\begin{center}
\includegraphics[width=7.0cm,clip]{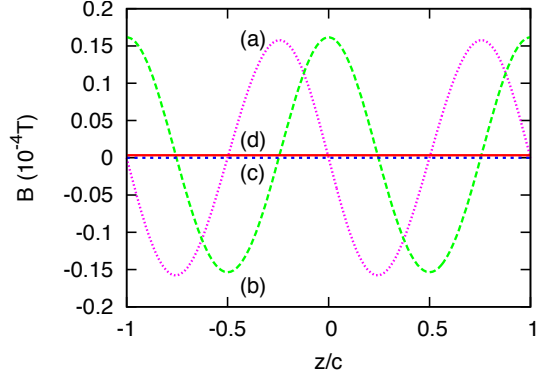}
\caption{(Color online) z dependence of $B_x$ and $B_z$ outside 
the superconductor for $T=0.185T_C$.  
(a) $B_z(x=-0.2c,z)$, (b) $B_x(x=-0.2c,z)$, 
(c) $B_z(x=-c,z)\times10$, and (d) $B_x(x=-c,z)\times10$.}
\end{center}
\end{figure}

Next we calculate the LDOS by employing  the prescription 
described in Ref. 13.  (In numerical calculations, we replace the
 $\delta$ function by a Lorentzian with the width $0.005J$.)
In Fig.6, the LODS at the surface site is shown for several 
temperatures.
At $T = 0.19T_c$, where Im $\chi = 0$, 
a zero-energy peak due to Andreev bound state appears. 
For $T < 0.19T_C$, flux phase order occurs and then  
the splitting of the peak takes place. 
When the temperature is decreased,  
Im $\chi$ becomes larger, and the splitting broadens 
as $T \to 0$. 
The theoretical result for the splitting at 
$T=0.01T_C$ ($\sim 1$K)
is large in comparison to the experiment.\cite{Coving}  
This discrepancy could be due to the neglect of 
fluctuations around the mean-field solution. 
Also, for the comparison to the  experiment, 
it should be noted that the LDOS at the surface directly corresponds 
to the tunneling conductance only when the interface is flat and clean 
and the transparency is low.\cite{5a,5b}

\begin{figure}[htb]
\begin{center}
\includegraphics[width=7.0cm,clip]{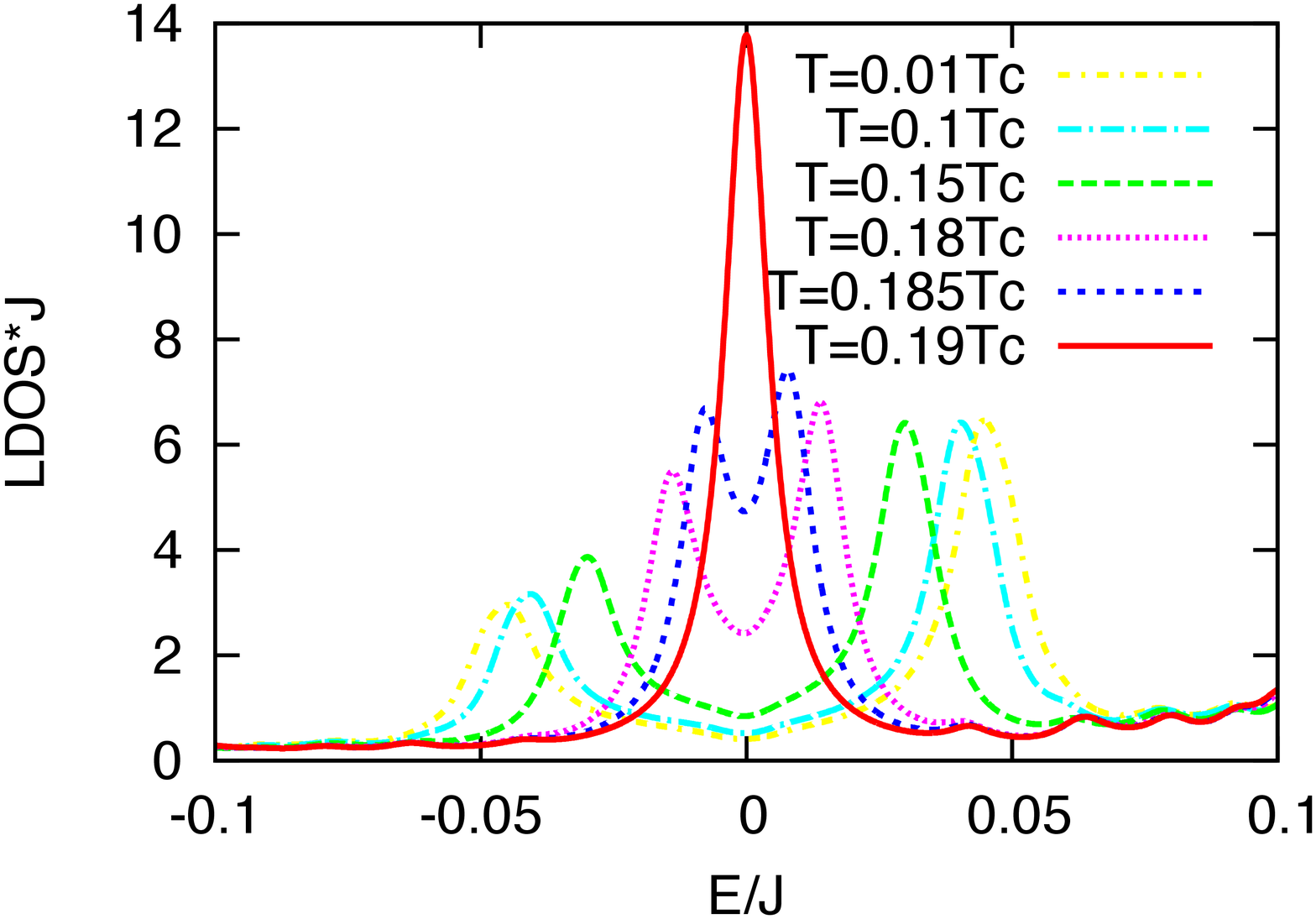}
\caption{(Color online) LDOS at the (110) surface for several 
temperatures. }
\end{center}
\end{figure}

In summary, we have studied the spontaneous magnetic 
field and the LDOS near the (110) surface of YBCO, 
where the time-reversal symmetry is spontaneously broken,  
by using the BdG method applied to the bilayer $t-J$ model.
It is found that in the superconductor spontaneous magnetic fields
can arise, but only in a narrow region   
with a width of the order of $\xi_d$ from the surface.
Outside the sample, the region in which $B$ is finite 
is much narrower; $B$ decays within a few \AA \
from the surface.  
These magnetic fields may be difficult to detect experimentally, 
but we expect $\mu$SR, polarized neutron scattering, and 
the spin polarized STM could detect them. 
We have also investigated the temperature dependence of the 
LDOS at the surface, and found that the spitting of 
the zero-energy peak decreases as the temperature is increased. 
 
For the values of $B$, our estimates give the upper bound, 
because we did not take into account the screening effect in 
superconductors (due to the use of Biot-Savart law), 
surface roughness, and the fluctuations around the mean-field 
solution. These effects will reduce not only the flux phase OP 
and the values of $B$, but also the splitting of the peak in the LDOS.
The latter could make the theoretical results get closer to the 
experimental values. 
 
There have been many other mechanisms aimed to 
explain the splitting of zero-energy peak observed by 
Covington et al.\cite{Coving} 
Effects of impurity scattering,\cite{6}
coexistence of spin-density wave\cite{Honer,8} and $d$-density 
wave order\cite{8} with $d$-wave SC states near the surface 
have been proposed. 
Also, $p$-wave SC order near the surface has been 
predicted.\cite{Honer,7b}
Since a large zero-energy DOS due to Andreev bound state exists 
near  a (110) surface of $d$-wave superconductor, 
the  system has a tendency toward ferromagnetism if  
an on-site repulsive interaction is present.\cite{Lee2,7a} 
This can lead to  edge ferromagnetic order with split peak. 

In the present work, the BdG equations are derived from 
the microscopic model and the parameters are chosen to 
represent the electronic structure of YBCO. 
This is the difference from other theories 
that have been proposed to explain both the apparent absence of 
spontaneous magnetic fields and the splitting of the zero-energy 
peak in the LDOS.
The results for the distribution of spontaneous magnetic fields 
calculated in this work may be tested experimentally, and 
it will decide whether the present theory correctly 
describes the surface state of YBCO.

\medskip
\begin{acknowledgment}
 The author thanks M. Hayashi and H. Yamase for useful discussions. 
\end{acknowledgment}


\end{document}